\def\etal{{\it et al.\/}}

\def\ie{{\it i.e.\/}}

\documentstyle[aps,prl,floats]{revtex}
\begin{document}
\draft
\tighten
\twocolumn[\hsize\textwidth\columnwidth\hsize\csname@twocolumnfalse%
\endcsname
\title{Ultra high energy neutrinos from gamma ray bursts}
\author{Mario Vietri}
\address{Dipartimento di Fisica E. Amaldi, Universit\`a di Roma 3
\\ via della Vasca Navale 84, 00147 Roma 
\\ E-mail: vietri@corelli.fis.uniroma3.it}
\date{\today}
\maketitle
\begin{abstract}
Protons accelerated to high energies in the relativistic shocks that generate
gamma ray bursts photoproduce pions, and then neutrinos {\it in situ}. 
I show that ultra high energy neutrinos ($> 10^{19}\; eV$) are produced 
during the burst and the afterglow. A larger flux, also from bursts, is 
generated via photoproduction off CMBR photons in flight but is not correlated 
with currently observable bursts, appearing as a bright background.
Adiabatic/synchrotron losses from protons/pions/muons are negligible.
Temporal and directional coincidences with bursts detected by 
satellites can separate correlated neutrinos from the background.
\end{abstract}

]

\narrowtext
The recent discovery of gamma ray bursts' (GRBs) afterglows \cite{obs},
accurately predicted by theoretical models \cite{theory} and disappearance 
of flares in the radio flux \cite{goodman} have bolstered our confidence 
in the correctness of the fireball model \cite{reviews}. According to 
the model, bursts are generated when two or
more hyperrelativistic shells, issued by an as yet unspecified source,
collide with each other. A relativistic shock forms, where non--thermal 
electrons are accelerated and then dissipate their internal energy through 
synchrotron (and possibly Inverse Compton) radiation. After the internal 
collision, the resulting shell will collide with the interstellar medium (ISM), 
thereby forming a second, relativistic shock, which will continue to 
expand into the ISM even after the burst proper, thusly generating the
afterglow.

The relativistic environment surrounding the above--mentioned shock
is ideal for the acceleration of protons to high energies \cite{lieu}.
The highest energy that can be attained is \cite{vietri}
\begin{equation}
\epsilon_{max} = 10^{20}\; eV \; \theta^{-5/3} \eta_2^{1/3} E_{52}^{1/3}
n_1^{1/6}\;.
\end{equation}
Here the explosion energy is $E = E_{52} 10^{52}\; erg$, the expansion
Lorenz factor $\eta = \eta_2 10^2$, the beaming angle $\theta$, and the
ISM number density $ n = n_1 \; cm^{-3}$. Currently popular values inferred
from afterglows are $\theta \approx 1/3$, $E_{52} \approx 1$ \cite{kulkarni}
implying $\epsilon_{max} \approx 6\times 10^{20}\; eV$.

When energetic protons interact with synchrotron photons emitted by electrons, 
they can produce pions; the decay of charged pions then produces electron 
and muon neutrinos. In this {\it Letter}, I will consider only ultra high 
energy neutrinos (UHENs, $> 10^{19}\; eV$) and neglect lower energy ones
\cite{bohdan,bahcall}.

\paragraph*{Expected fluxes of Ultra High Energy Neutrinos.}

Let us consider a burst of duration $T$ seconds; according to the fireball 
theory for external shocks \cite{reviews}, this occurs at a distance $r_e = 
\eta^2 c T$ from
the unspecified burst source, and, in the shell frame, the shell thickness is
$\delta\!r = \eta c T$. The total energy density
in the shell frame is then $U_\gamma = L_\gamma/4\pi r_e^2 c \eta^2$; inserting
this into Eq. 3 of ref. \cite{bahcall} I find the inverse of the
timescale for photopion losses, $t_\pi^{-1}$; multiplying by the time the 
proton spends in the shell, in the shell frame ($= \eta T$), I find that, for a 
proton of energy $\epsilon_p$ as seen by an outside observer, immersed in a 
radiation field with turnover frequency $\epsilon_\gamma \approx 1\; MeV$, 
beyond which the spectrum significantly steepens, the total probability for 
photopion production is 
\begin{equation}
\label{prob}
f_\pi^{(0)} = 0.03 \eta_2^{-4}
\frac{L_\gamma}{10^{50}\; erg\; s^{-1}}
\frac{1\; MeV}{\epsilon_\gamma}
\frac{10\; s}{T}\;,
\end{equation}
for proton energies exceeding \cite{bahcall}
$\epsilon_b = 2\times 10^{15}\; eV \eta_2^2 (1\; MeV/\epsilon_\gamma)$.
I have used here a typical luminosity for long--lasting bursts, such as those
with the ISM are thought to be, and a typical long duration. 

Experiments such as AIRWATCH \cite{linsley,takahashi} have appreciable detection
efficiencies for neutrinos exceeding the threshold energy 
$\epsilon_{\nu,l}\approx 10^{19}\; eV$. Since neutrinos emitted through 
photopion processes typically carry away a fraction $q \approx 0.05$ of the 
proton energy (losses will be discussed later), I have to
compute the energy release in protons with energies exceeding $\epsilon_l =
\epsilon_{\nu,l}/q \approx 2\times 10^{20} \; eV$. The spectrum in high energy 
protons accelerated at relativistic shocks is roughly $\propto \epsilon^{-2}$, 
and defining the total energy released in ultra high energy cosmic rays (UHECRs,
$\epsilon > \epsilon_1 = 10^{19}\; eV$) as  $E_U$, I have that the 
whole energy in UHECRs which can emit detectable UHENs ({\it i.e.}, 
$\epsilon > \epsilon_l =\epsilon_{\nu,l}/q$) is $E_U \ln 
\epsilon_{max}/\epsilon_l / \ln \epsilon_{max}/\epsilon_1$. 
Only a fraction $2 q f_\pi^{(0)}$ of this ends up in UHENs. 
Thus the total energy emitted in UHENs is 
\begin{equation}
E_\nu = 2 q f_\pi^{(0)} E_U \frac{\ln \epsilon_{max}/\epsilon_l}
{\ln \epsilon_{max}/ \epsilon_1} \;.
\end{equation}
The total flux 
of UHENs can then be obtained by integrating the flux over all distances:
\begin{equation}
\dot{n}_\nu = \dot{n}_{GRB} \frac{E_\nu}{\bar{\epsilon}_\nu} \frac{c K}{H_\circ}
= 2 q f_\pi^{(0)}\frac{\dot{n}_{GRB} E_U}{\bar{\epsilon}_{\nu}
\ln \epsilon_{max}/\epsilon_1}
\frac{c K}{H_\circ}\;,
\end{equation}
where $\bar{\epsilon}_\nu = \epsilon_{\nu,l} \ln \epsilon_{max}/\epsilon_l$ 
is the average neutrino energy from this process, and the delicate factor $K$, 
to be discussed later on, 
takes account of such unknowns as the GRBs' redshift and luminosity 
distributions, and the details of the cosmological model. 

The dependence of these neutrino rates upon physical factors of individual
bursts, such as $\eta, L_\gamma,$ 
and $\epsilon_\gamma$ is all contained within $f_\pi^{(0)}$ (Eq. 
\ref{prob}), and will be omitted from now on for sake of conciseness. 
The key factor in the above equation is $\dot{E} = \dot{n}_{GRB} 
E_U$, the injection rate per unit volume of non--thermal proton energy,
because the others either are known or enter logarithmically. 
It is known already that, under the hypothesis that GRBs emit about as much 
energy in the form of $\gamma$--band photons and UHECRs, the flux of UHECRs 
at Earth is reproduced to within a factor of $\approx 3$ \cite{vietri,wax}.
I will show later that UHECRs are accelerated within afterglows, which 
dominate the energy balance by about a factor of $10$. Then, if the same
rough equipartition between radiation and UHECRs holds during the afterglow,
the total energy release required to explain the UHECR seen at Earth is
correctly accounted for. 

That the equipartition argument yields a correct answer can be checked by 
considering that the observed burst rate ($\approx 30\; yr^{-1}\; Gpc^{-3}$) 
times the observed energy release including afterglow ($\approx 10^{52}\; erg$) 
yields an energy release rate, $3\times 10^{44}\; erg\; yr^{-1}\; Mpc^{-3}$, 
very close to that deduced \cite{wax2} without explicit reference to the
nature of the sources of UHECRs: $\dot{E} = 4.5\times 10^{44}\; erg\; yr^{-1}\; 
Mpc^{-3}$ for the restricted range of proton energies $10^{19}\; eV < \epsilon 
< 10^{21}\; eV$. 

Thus, under the equipartition assumption I can use the energy release necessary 
to explain Earth observations as the energy released in UHECRs by GRBs; 
taking $\epsilon_1 = 10^{19}\; eV$, and defining 
$H_\circ \equiv h 50 \; km\;s^{-1}\; Mpc^{-1}$, I obtain
\begin{equation}
\label{flux}
\dot{n}_\nu = 2.2\times 10^{-11} \frac{f_\pi^{(0)}}{0.03} h^{-1} K 
yr^{-1} \; cm^{-2}\;.
\end{equation}

The flux determined above is {\it not} the whole flux of UHENs from GRBs 
detectable at Earth. The reason is that {\it all} UHECRs eventually will
emit UHENs by photoproduction with photons of the CMBR, the so--called
Greisen--Zatsepin--Kuz'min effect \cite{gzk}. This neutrino production
will occur in flight, rather than {\it in situ}, with a typical mean
free path of order $\approx 10 Mpc$. As they cross this distance, UHECRs
are slowed down in their progress toward Earth by the turbulent intergalactic 
magnetic field. While estimates of this delay are very uncertain because of our 
ignorance of both strength and correlation length of the field, they still
all agree in putting it above $10^2-10^3\; yr$, \ie\/ in washing away any
correlation with GRBs observed within our lifetimes. The total
flux of background UHENs $\dot{n}^{(bg)}_\nu$, uncorrelated with observable
GRBs, is thus 
\begin{equation}
\label{background}
\dot{n}^{(bg)}_\nu = \frac{\dot{n}_\nu}{f_\pi^{(0)}} = 7.3\times 10^{-10} K 
h^{-1} yr^{-1} \; cm^{-2}\;.
\end{equation}

The computation of the factor $K$ requires an explicit hypothesis on the
distribution of redshifts and luminosities of GRBs. A detailed computation
\cite{yoshida} for idealized redshift distributions of standard candles, has 
been carried out. Comparison of Table 1 of ref \cite{yoshida} with
the above equation shows that their computed values of $K$ vary by a 
factor of $3$ either side of the value I obtained. 

\paragraph*{Afterglows.}

I show now that acceleration of protons to the highest energies does continue 
unabated through most of the afterglow. After the burst, the relativistic
shell keeps plowing through the interstellar medium, sweeping up more matter
and decelerating. The shell Lorenz factor scales as
$\eta = 6.4 n_1^{-1/8} E_{52}^{1/8} t_d^{-s}$, where $t_d$ is the post--burst
time in days neglecting redshift. For adiabatic expansion $s=3/8$ 
\cite{kulkarni} while $s=3/7$ for radiative expansion \cite{vietrigrb}.
The maximum energy of non--thermal protons (Eq. 1) decreases very slowly with 
time, as $t^{-1/8}$ or $t^{-1/7}$ for adiabatic or radiative expansion, 
respectively. In particular, for the best values $E_{52} = 1$ and $\theta =
1/3$, production of UHENs ceases (\ie\/, $\epsilon_{max}
< \epsilon_{\nu.l}/q$) for $\eta < 3.3$, corresponding to $\approx 6\; d$ after 
the burst, nearly independent of whether expansion is adiabatic or radiative.

I also have to check that the probability of photopion production through
the afterglow does not change by much from the value computed (Eq. \ref{prob})
for the burst proper. This requires some discussion.
From observations \cite{fruchter} we know that
the instantaneous luminosity scales as $t^{-\alpha}$, with $\alpha \approx
1.1$. Also, we know from fireball theory that $T \propto
r/\eta^2$, and that  $r \propto \eta^{-v}$, where $v= 2/3$ for adiabatic
or $v=1/3$ for radiative expansion. 
So the factor $L_\gamma \eta^{-4} T^{-1} \propto t^{s(2-v)-\alpha}$. However,
it is more difficult to establish the variation of the spectral break
$\epsilon_\gamma$ with time, which is not currently observed. It seems 
however that, given the general softening of radiation within the afterglow,
it is unlikely to remain constant; a more likely hypothesis 
is that it decreases slowly with time. Phenomenologically, one may take
$\epsilon_\gamma  \propto \eta^q$.
The limits within which $q$ is expected to vary are easy to ascertain. On
the one hand, $q =0$ would imply that the cut--off does not evolve, despite
the shell slow--down. This is both unphysical, and contrary to some weak
evidence that it may decrease within the burst proper. On the other hand,
the synchrotron turn--on frequency (\ie, that beyond which synchrotron 
emits most of the energy) scales as $\propto \gamma^4$; in the afterglow model,
all emission is due to synchrotron processes. However, the very long lasting 
optical emission from GRB 970228 seems to imply a very 
extended synchrotron spectrum, so that $q=4$ may be considered an upper limit. 
Thus $0 < q < 4$. I then obtain $\epsilon_\gamma \propto \eta^q \propto 
t^{-qs}$. From Eq. \ref{prob} I then find $f_\pi^{(0)} \propto t^z$, with 
$z= s(q+2-v)-\alpha$. Only taking a small value, $q=1$, and then only for 
adiabatic expansion, do I find $z < 0$. Thus we see that overall, the
probability $f_\pi^{(0)}$ is unlikely to decrease: if anything, $f_\pi^{(0)}$
is likely to {\it increase} through the afterglow, so that our estimates 
are, most likely, lower limits. Thus, by taking in the previous section 
$f_\pi^{(0)} \approx $ constant, I did not overestimate the neutrino fluxes.
An interesting consequence of this is that the luminosity in UHENs scales 
approximately as $L_\nu = f_\pi^{(0)} L_\gamma \propto t^{-1}$, which means 
that equal logarithmic post--burst--time intervals are equally likely to 
contain an observable neutrino. 

\paragraph*{Losses.}

Proton losses (synchrotron and photohadronic) were shown to be negligible
in ref. \cite{vietri}: the proton energy is limited by the size of the shell. 
I have to consider however adiabatic and synchrotron losses by pions and 
muons, which could considerably limit the highest energies achieved by 
neutrinos.

Adiabatic losses are significant whenever the particle lifetime 
$\gamma_\star\tau_\star$ 
in the shell frame exceeds the characteristic timescale on which the 
magnetic field decreases because of the shell expansion; here $\star$ indicates
either pion or muon, and $\tau_\pi = 2.6\times 10^{-8}\; s$ and 
$\tau_\mu = 2.2\times 10^{-6}\; s$ are their respective lifetimes in their 
rest frames. The limiting Lorenz factors are found when the two timescales
match, \ie, when $\gamma_\star\tau_\star = 2B/\dot{B}$. Following ref. 
\cite{rachen}  I shall take $B \propto R^{-2}$, where $R$ is the 
transverse dimension of the causally connected region, which, following
refs. \cite{rachen,laguna}, is
given by $r/\eta$, even through the afterglow, and obviously $\dot{R} \approx 
c$. Then I obtain the limiting Lorenz factor in the observer frame $\gamma_l = 
r/c \tau_\star$, independent of whether the afterglow is adiabatic or radiative.
Scaling $r\equiv x r_i$ by its lowest value, that at the moment of the burst
proper, $r_i = 2\eta^2 cT = 6\times 10^{15}\; cm \eta_2^2 (T/10\;s)$,
I find $\gamma_\pi = 10^{13} 
x$, and  $\gamma_\mu = 10^{11} x$ for $x \geq 1$, both exceeding the 
proton's Lorenz factor in Eq. 1. For protons with Lorenz factor $\gamma_p$ in 
the shell frame, the synchrotron cooling time is $t_s = 1 \; yr\; (10^{11}/
\gamma_p) (1 \; G/B)^2$. For synchrotron losses to be negligible, the Lorenz
factor of pions/muons must not exceed the limiting $\gamma_\star$ given by 
\cite{rachen} $\gamma_p t _s 
(m_\star/m_p)^3 = \gamma_\star^2 \tau_\star$, where $m_\star/m_p \approx 0.1$
for both pions and muons. From ref. \cite{laguna}, $B \approx 
1\; G\; \eta_2^{1/2}$ for the external shock scenario and the
afterglow. Transforming back to the observer frame I find $\gamma_\pi = 
3\times 10^{13} \eta_2^{1/2}$, and $\gamma_\mu = 3\times 10^{12} \eta_2^{1/2}$. 
Both exceed the Lorenz factors of the proton, Eq. 1. Thus 
adiabatic/synchrotron losses of pions/muons do not affect the arguments of
this paper. 

\paragraph*{Detectability.}

Currently planned experiments such as AIRWATCH \cite{takahashi} will 
monitor from satellites fluorescent light profiles of cosmic ray cascades over 
areas of order $A = A_6 \times 10^6\; km^2$, with $A_6 \approx 1$. 
The interaction probability for UHENs is 
proportional to the monitored column density ($10^3\;g \; cm^{-2}$); it also 
depends over the extrapolation of the cross--section to currently unobserved 
energies, but typical values are $\sigma \approx 3\times 10^{-32} cm^2\;
(\epsilon_\nu/10^{19}\; eV)^{1/2}$ 
\cite{takahashi,quigg}. Once the neutrino has interacted, a detection 
efficiency close to $1$ for UHENs is reported by feasibility studies, at
energies $\epsilon_\nu \approx 10^{19}\; eV$,  
yielding interaction probabilities of $P_\nu  \approx 3\times 10^{-5}$. 
This translates into an expected number of detectable UHENs of
\begin{equation}
\label{obs}
\dot{N}_\nu = 7\; K \; A_6 \; yr^{-1} \frac{f_\pi^{(0)}}{0.03} h^{-1} \;.
\end{equation}
At the same time, we expect a background flux from Eq. \ref{background} given by
\begin{equation}
\label{back}
\dot{N}_\nu^{(bg)} = 200 \; K\; A_6 \; yr^{-1} h^{-1}\;.
\end{equation}
It is safe to state that Eqs. \ref{obs} and \ref{back} have large errors,
due to our ignorance both of the neutrino--nucleon cross--section at these
large, and untested neutrino energies, and to the sources' redshift
distribution (the parameter $K$). 

The requirement that the expected number of neutrinos correlated with bursts be 
large enough to ensure detection within a year of operation can be turned, using
Eq.  \ref{prob}, into a requirement on the area covered by the experiment:
\begin{equation}
\label{condition}
A_6 \gg 0.2 K^{-1} \eta_2^4 \frac{10^{50}\; erg\; s^{-1}}{L_\gamma}
\frac{\epsilon_\gamma}{1\; MeV} \frac{T}{10\; s}\;.
\end{equation}
Detection of correlated neutrinos seems possible provided bursts due external 
shocks are well--represented by the average values employed above. 

The flux of Eq. \ref{back} of an event per day, completely uncorrelated with 
currently observable 
bursts, obliges us to face the issue whether we can distinguish from casual 
associations a much smaller ($f^{(0)}_\pi \approx 0.03$) flux which is indeed 
correlated (to within the afterglow duration, $\approx 6\; d$)
with simultaneously observed bursts. The answer would be an easy yes
if UHENs arrived simultaneously with the burst proper, because we could
then use very tight directional and temporal coincidences to distinguish 
the signal from background noise. But, since I argued above that most
neutrinos are produced during the afterglow which is observed to last 
for a few days after the burst, it has to be ascertained whether this can
still be done. The answer is a qualified yes. 

Suppose I can measure the directions of arrival of neutrinos and GRBs
with a combined directional error of $\beta$. Calling $\dot{N}_{GRB}$ the rate 
of detection of GRBs in the $\gamma$--ray, the probability of casual 
association $P_c$ is 
\begin{equation}
\label{simul}
P_c = \dot{N}_{GRB} \delta\!t \frac{\beta^2}{4} = 4\times 10^{-4}
\frac{\delta\!t}{6\; d}\left(\frac{\beta}{1^\circ}\right)^2
\end{equation}
where I used $\dot{N}_{GRB} = 300\; yr^{-1}$, typical of BATSE \cite{fishman}.
The rate of appearance of casual associations is $ \dot{N}_\nu^{(bg)} P_c
\approx 0.08 \; yr^{-1}$, reassuringly smaller than the rate of
physical associations, Eq. \ref{obs}. This condition, $P_c \dot{N}_\nu
^{(bg)} \ll \dot{N}_\nu$, can also be written as
\begin{equation}
\label{feasible}
f_\pi^{(0)} \gg \dot{N}_{GRB} \delta\!t \frac{\beta^2}{4} \approx 4\times 
10^{-4} \frac{\delta\!t}{6\; d} \left(\frac{\beta}{1^\circ}\right)^2 \;.
\end{equation}
Comparison with Eq. \ref{prob} shows that the experiment can be done, 
provided angular errors of order 
\begin{equation}
\label{resolution}
\beta \ll 7^\circ \left(
\frac{L_\gamma}{ 10^{50} \; erg\; \eta_2^{4}} 
\frac{\epsilon_\gamma} {1\; MeV} \frac{T}{10\; s}
\frac{\delta\!t}{6d}
\right)^{1/2}
\end{equation}
can be achieved. 

Lastly, since the rate of Eq. \ref{back} is comparable to that of
GRBs detected by BATSE \cite{fishman}, measurement of dipole
and quadrupole moments of the neutrino distribution may just be doable. 
The spectrum of UHENs (both background and correlated ones) will follow 
accurately that of UHECRs in GRBs, since the probability of photopion losses 
(Eq. \ref{prob}) is independent of proton energy. It should thus be possible 
to see the cutoff in the UHECR spectrum, Eq. 1, as mirrored in neutrinos. 

\paragraph*{Discussion.}

The acceleration of UHECRs in GRBs is so effective, that it has been proposed
\cite{vietri,wax}
that the whole flux of UHECRs at Earth comes from these
events. However, since UHECRs can take $\approx 10^3\; yr$ more than photons
to reach us from the closest GRBs, it will be impossible, within
our finite lifespans, to establish a direct association between GRBs and UHECRs.
A sure hint should be that no AGNs, or peculiar object, ought to be seen
close to the direction of arrivals of UHECRs, but this expectation is not
unique to this model, and is common for instance to strings. On the other 
hand, a UHEN of $\approx 10^{19}\; eV$ would accumulate with respect to photons 
emitted simultaneously a delay of only $\approx 10^{-19}\; s
(m_\nu/10\; eV)$ in coming from even a distance of $c/H_\circ$, the radius of 
the Universe, with $m_\nu$ the neutrino mass. Thus it would be 
essentially simultaneous to photons (including afterglow's photons).
Furthermore, the UHENs can only be 
produced by the highest energy protons, those, in other words, well beyond the 
Greisen--Zat'sepin--Kuzmin limit. Thus the UHENs produced {\it in situ}
represent the surest smoking gun that UHECRs are accelerated in GRBs. 
Different, electromagnetic signatures of the association of UHECRs and GRBs 
have been discussed in refs. \cite{gev,dermer}.

In short, what detection of UHENs will allow us to do is to circumvent the
shortsightedness imposed upon us by the Greisen--Zatsepin--Kuz'min limit,
and to investigate the generation of the highest energy cosmic rays throughout
the whole Universe. 
The only alternative sources of UHENs proposed so far are cosmic strings
\cite{strings} and AGNs \cite{biermann} which are also the only alternative 
sources proposed so far for UHECRs. I have discussed here that a fraction of 
all UHENs (Eq. \ref{prob}) should show an association with simultaneously 
observed GRBs, if they indeed originate in GRBs. Thus a potentially clear--cut 
way to distinguish between the three competing theories is available
and it might, perhaps, already be accessible to AIRWATCH--class experiments.

Thanks are due to L. Scarsi and J. Rachen for helpful comments.


\begin{references}
\bibitem{obs} In the X--ray, Costa, E. \etal, 1997, Nature, {\bf 387}, 783;
in the optical, van Paradijs, J., \etal, 1997, Nature, {\bf 386}, 686;
in the radio, Frail, D.A., \etal, 1997, Nature, {\bf 389}, 261.
\bibitem{theory} In the radio, B. Paczy\'nski, and J.E. Rhoads, Ap.J.L., 
{\bf 418}, L5 (1993); in the optical, P. M\'esz\'aros, and M.J. Rees, Ap.J., 
{\bf 476}, 232 (1997); in the X--ray, M. Vietri, Ap.J.L., {\bf 478}, L9 (1997).
\bibitem{goodman} J. Goodman, New Astronomy, {\bf 2}, 449 (1997).
\bibitem{reviews} For reviews, see T. Piran, in {\it Some unsolved problems in
Astrophysics}, Princeton Univ. Press, Princeton 1995, eds. J. Bahcall and 
J.P.Ostriker, p. 435; M.J. Rees, astro--ph n. 9701162 (1997).
\bibitem{lieu} J.J. Quenby, Lieu, R., Nature, 342, 654 (1989).
\bibitem{vietri} M. Vietri, Ap.J. {\bf 453}, 883 (1995);
the dependence of Eq. 1 upon $n_1$ differs from that in
this reference because a small error has been corrected. 
\bibitem{kulkarni} E. Waxman, S. Kulkarni, D.A. Frail, submitted to
Ap.J.L., astro-ph. n. 9709199 (1997).
\bibitem{bohdan} B. Paczy\'nski and G. Xu, Ap. J., {\bf 427}, 708 (1993).
\bibitem{bahcall} E. Waxman and J. Bahcall, Phys. Rev. Lett., {\bf 78}, 2292
(1997).
\bibitem{linsley} J. Linsley, in {\it Proc. of the XXVth ICRC}, Durban RSA,
{\bf 5}, 381 (1997).
\bibitem{takahashi} Y. Takahashi, \etal, SPIE 2806-13 (1996); J. Linsley \etal, 
in {\it Proc. of the XXVth ICRC}, Durban RSA, {\bf 5}, 385 (1997); P. 
Attin\`a \etal, {\it ibidem}, {\bf 5}, 389 (1997); 
J. Forbes \etal, {\it ibidem}, {\bf 5}, 273 (1997).
\bibitem{wax} E. Waxman, Phys. Rev. Lett., {\bf 75}, 386 (1995).
\bibitem{wax2} E. Waxman, Ap. J. L. {\bf 452}, L1 (1995).
\bibitem{gzk} K. Greisen, Phys. Rev. Lett.
{\bf 16}, 748 (1966); G.T. Zatsepin and V.A. Kuz'min, JETP Lett. {\bf 4},
78 (1966); R.J. Protheroe and P.A. Johnson , Astroparticle Phys. {\bf 4},
253 (1995) and {\it erratum} {\bf 5}, 215 (1996).
\bibitem{yoshida} S. Yoshida and M. Teshima, Prog. Theor. Phys., {\bf 89}, 
833 (1993).
\bibitem{maoz} E. Maoz, 1994, ApJ, {\bf 428}, 454 (1994).
\bibitem{vietrigrb} M. Vietri, Ap.J.L., {\bf 488}, L105 (1997).
\bibitem{fruchter} A. Fruchter {\it et al.}, IAU Circ. n. 6747 (1997).
\bibitem{rachen} J.P. Rachen, and P. M\'esz\'aros, submitted to Phys. Rev. D,
astro--ph. n. 9802280 (1998).
\bibitem{laguna} P. M\'esz\'aros,  P. Laguna, and M.J. Rees, ApJ, {\bf 405}, 
278 (1993).
\bibitem{quigg} C. Quigg, M.H. Reno, and P.T. Walker, Phys. Rev. Lett., 
{\bf 57}, 1204 (1986).
\bibitem{fishman} G.J. Fishman and C.A. Meegan, ARAA, {\bf 33}, 415 (1995).
\bibitem{gev} M. Vietri, Phys. Rev. Lett., {\bf 78}, 1323 (1997). 
\bibitem{dermer} M. Boettcher and C.D. Dermer, sumitted to ApJ, astro--ph. n.
9801027 (1998).
\bibitem{strings} G. Sigl, S. Lee, D. Schramm and P. Bhattacharjee, Science,
{\bf 270}, 1977 (1995).
\bibitem{biermann} J.P. Rachen and P.L. Biermann, Astron. Ap., {\bf 272}, 
161 (1993).
\end{references}
\end{document}